# Ecosystem-Agnostic Standardization of Quantum Runtime Architecture: Accelerating Utility in Quantum Computing.

1st Markiian Tsymbalista
*Faculty of Electronics and Computer Technologies*
*Ivan Franko National University of Lviv*
Lviv, Ukraine
markiian.tsymbalista@lnu.edu.ua

2nd Ihor Katernyak
*Faculty of Electronics and Computer Technologies*
*Ivan Franko National University of Lviv*
Lviv, Ukraine
ihor.katernyak@lnu.edu.ua

*Abstract*—Fault tolerance is a long-term objective driving many companies and research organizations to compete in making current, imperfect quantum computers useful - Quantum Utility (QU). It looks promising to achieve this by leveraging software optimization approaches primarily driven by AI techniques. This aggressive research covers all layers of Quantum Computing Optimization Middleware (QCOM) and requires execution on real quantum hardware (QH). Due to the nascent nature of the technology domain and the proprietary strategies of both large and small players, popular runtimes for executing quantum workloads lack flexibility in programming models, scheduling, and hardware access patterns, including queuing, which creates roadblocks for researchers and slows innovation. These problems are further exacerbated by emerging hybrid operating models that place Graphical Processing Unit (GPU) supercomputing and Quantum Intermediate Representation (QIR) at the heart of real-time computations across quantum and distributed resources. There is a need for a widely adopted runtime platform (RP) driven by the open-source community that can be easily deployed to work in a distributed manner between Quantum Processing Unit (QPU), GPU, control hardware, external compute resources and provide required flexibility in terms of programming & configuration models.

## Introduction

QCOM [1] as complex optimization software layer will continue it is evolution driven by an effort of significant number of researchers. While some of the layers in QCOM don't require efficient combination of classical and quantum computations and remote execution, for effective computation optimization techniques it is mandatory. There are several obstacles to that mostly driven by the fact that access to QH is scarce and is mostly exposed through proxy companies building their own runtimes which usually implies a lot of limitations of how APIs could be used. They are tuned for regular quantum algorithms development, lack flexibility and hide implementation details under the hood. By doing this they are reserving their part of the pie. But successful research requires uncertainty and black box points to be removed to be able to find gaps in the value stream and prepare proper compensation strategies. Whether it is quantum or well know classical – it is still software. Every software requires it is builders to understand architecture, it is core qualities and limitations. So, having closed eco-systems on the path to QU creates a significant bottleneck. From another perspective open-source community for quantum software is growing fast. Open Quantum Hardware [3] initiative aims to put under umbrella all open-source quantum tech. But there is no mature project for RP which could be adjusted to personal needs of researchers and deployed close to QH where maximum flexibility could be leveraged to move things forward. Smaller industry players are spending a lot of time reinventing the wheel by implementing the same runtime capabilities which are said not to be a "rocket science" but certainly piece of tooling that takes several men months to build, not considering dependencies on the approval and deployment of those runtimes by hardware vendors.

There is an importance of building value on top and move forward which unfortunately doesn't happen to RPs nowadays. Back in the days in Richard Hamming in his famous lecture "You and Your Research" [2] stated the following: "These days we stand on each other's feet! You should do your job in such a fashion that others can build on top of it, so they will indeed say, "Yes, I've stood on so and so's shoulders and I saw further." The essence of science is cumulative." There is a need for a quantum runtime reference architecture that will be closing common challenges of researchers in the path to QU. It should be further implemented in working software. Adoption & development by community will play a significant role, while being open source in it is nature. Open source accelerates innovation and business value generation as has been seen with Linux and strategic transformation of .NET Framework from proprietary technology to open source one that has changed it and industry practices 10x into positive direction. Open-sourcing quantum RP should be significant milestone of driving eco-system forward.

## METHODOLOGY

Upon implementation of experiments as part of QCOM scope, authors encountered a list of challenges. These scenarios require system capabilities that are not widely implemented in existing tools and RPs. This has been observed as part of technology consulting effort for one of the commercial companies working in quantum optimization research. To refine product context, Value Proposition Canvas [5] has been leveraged. Further, by using system context elaboration techniques, use-cases and constraints were identified which served as an input to the software analyses and design effort. Software architecture methodology and it is methods including Quality Attributes Scenarios, Attribute Driven Design (ADD) [8] and Architecture Tradeoff Analyses Method (ATAM) [9] were leveraged to think of, design and evaluate a blueprint of a runtime reference architecture to cover required system capabilities.

Paper also utilizes methods from qualitative research to analyze past work that has been proposed on quantum cloud platforms and runtimes. Open Hardware Solutions in Quantum Technology [3] initiative provided a motivation to expand a family with a new open-source runtime reference architecture and solution in the future.

## VALUE PROPOSITION SUMMARIZED

Product offering:
- Software RP that allows to expose access to QH, mirroring core functionality of leading cloud quantum services (not including QH itself) and exposing additional features.
- Client Software Development Kit (SDK) that is leveraged by software clients to interact with remote runtime.
- Programming model based on adapter design to streamline common quantum performance optimization challenges.

Jobs to be done:
- Quantum Machine Learning (QML) execution. Efficiently run ansatz optimization loops (schedule a batch of independent jobs).
- Compilation. Optimize compilation flows for circuit parts that require frequent recompilation as an example.
- Noise mitigation: efficiently run pre- and post-process error mitigation (EM) schemes that require multiple modified quantum circuits (ZNE, PEC, Tomography-based methods, etc.).
- Measurements processing e.g. run advanced EM post-processing routines involving DNNs or Tensor-Networks.
- Combine QML + EM uses cases in different combinations
- Have a baseline solution to analyze and implement support for complex distributed use cases. Example could be to schedule a circuit, part of which is going to be run on the simulator (maybe even specific GPU hardware) and another part on real QH. Further, this use case could be split into two modes: with mid-circuit measurements and without.
- Batch job that includes different circuits with ability to choose specific EM method for each.
- Deploy custom library on the runtime for gates calibration, EM or other accuracy optimization task, so it could be used during computations.
- Schedule an arbitrary circuit implemented with common frameworks e.g. Qiskit, Cirq, PennyLane for execution on QH.

Gains that researchers are looking for in a platform:
- A need to put circuit execution closer to QH.
- Minimal impact of RP on results accuracy.
- Code reuse for tasks like EM, QML.
- Plugin resource estimation capability.
- Quick deployment on arbitrary compute infrastructure. Full infrastructure automation of the platform - Infrastructure as Code (IaC) approach.
- Familiar tech stack.
- Usability of software interfaces.
- Cost savings.
- Flexibility of programming model that allows full customization of execution pipeline.

Pains that researchers are having:
- For niche quantum hardware manufacturers and research institutions a necessity to build from scratch runtime/cloud quantum platforms to expose their QH.
- Long execution time for circuits. QML use cases spawn thousands of calls to QH when running optimization loops for ansatz. QH could be located on another continent. Because of that network operations could significantly increase waiting time for algorithm execution, making researchers to be idle like in the old days of mainframe machines.
- Low availability of QH for running workloads (mechanisms for queue management and resource allocation don't provide required, optimal level of determinism).
- Dependency on vendor SDKs (with qiskit runtime [4] dependency on qiskit) and their limited customization e.g. compilations steps, hardware access approach, opinionated runtimes primitives (with some extensibility points though).
- No optimal structure of runtime pricing models from different vendors.
- Black box architecture and implementation details of most quantum platforms.
- Lack of customization of RPs.

Gain creators
- Open-source solution that could be deployed to commodity classical compute infrastructure either public or private.
- Co-location deployment model reduces impact on results accuracy.

- Programming model that includes primitives required to solve common quantum tasks.
- Ability to plugin resource estimation module.
- Implemented using well-known tooling e.g. python, docker, kubernetes.
- SDK implemented adhering to object-oriented design practices.
- Costs savings by using open-source platform that doesn't require licensing costs.
- Flexibility of the programming model to construct and run complex custom pipelines.

Pain relievers
- Efficiency when running workloads that require thousands of circuit executions as part of one algorithm.
- Queue management and resource allocation modules that are opened for extensibility.
- Ability to use arbitrary quantum circuits framework.
- No dependency on pricing policies for bridge services e.g. AWS, Azure due to open-source nature.
- Documented architecture with open-sourced codebase.
- Flexibility to substitute or customize every component that constitutes platform.
- For small quantum hardware fabs ability to leverage open-source ready to use solution to expose access to their hardware and start earning revenue.

Existing RPs that are close to what is expected but still miss one or more points which is critical:
- Qiskit runtime [4]. Lack of flexibility in programming model, complex extensibility, dependance on IBM ecosystem, no details on internal design.
- Qiskit Dell Runtime [34]. Depends on a lot of outdated libraries, not actively supported, dependance on Qiskit, error-prone & hard to maintain programming model.
- Strangerworks qiskit runtime [35]. Building abstraction extending Qiskit runtime primitives that allows to access different back-ends (from different providers) in Qiskit programming model. Lack of flexibility to use other SDKs, proprietary technology.
- Pennylane Catalyst [36]. Part of the product contains runtime. Tied to the Pennylane eco-system and framework. Runtime implements some logic around device management which is not explicit as a result not extensible.
- Intel runtime [30]. As part of the SDK is a runtime. There is no much visibility of how it is built under the hood. What is obvious is that it's C++ first.
- Multiple papers e.g. [37], [38] on RPs are addressing some of the related challenges, but don't cover value proposition outlined above.

CONCEPTUAL ARCHITECTURE VISION FOR QCPAAS

Core blueprint of the Quantum Computing Platform as a Service (QCPaaS) architecture is outlined on the Fig.1. Next paragraph describes architecture using the concept of Architecture Decision (AD) which is part of ADD [8] architecture methodology in a narrative style.

It is assumed to be built around microservices reference architecture (AD-1), heavily reliant on docker as a containerization technology and on Kubernetes as a cluster orchestration platform (AD-2). Platform itself is scripted with IaC approach using Terraform [6] software tool (AD-3) what allows to deploy platform in either private or public data centers with couple of clicks and with minimal manual configuration. Platform concept of workers implement Bring Your Own Container (BYOC) pattern what allows to extend core image of the worker to include any custom library e.g. Mitiq [7] (AD-4). It means that in a working condition several workers with required library images are deployed to the cluster and are used by end users of the QCPaaS by passing an option in the SDK call. No explicit AD about Continuous Integration/Continuous Deployment (CI/CD) server is done but BYOC is deployed to the cluster by using the pipeline job of the CI/CD server by triggering it manually or using platform User Interface (UI) that could be additionally developed (AD-5).

Description of the main components and their relationships is defined below:

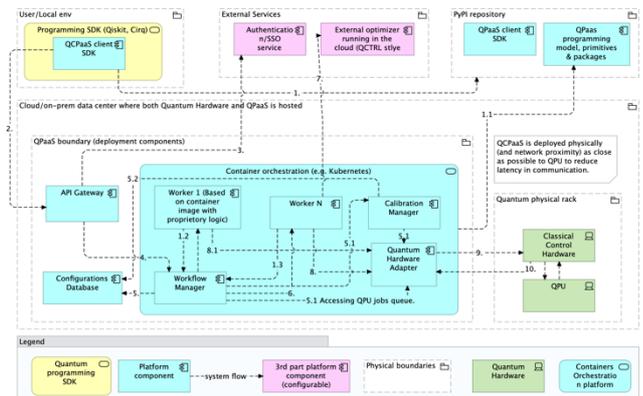

Fig. 1. Conceptual architecture QCPaaS

1. User installs SDKs for communication with QCPaaS (contains programming model).
    1.1. Containers running on cluster access QCPaaS libraries if needed.
    1.2. Worker 1 registers itself with scheduler, sharing details about compute capability.
    1.3. Worker N registers itself with the scheduler.
2. User creates quantum & classical program using framework of choice, schedules it for execution on the QCPaaS (leveraging platform SDK).
3. Client is authenticated with external service. This is a configurable element, particular implementation of which could be substituted. There is also no explicit choice regarding API gateway implementation.
4. Workflow Manager component receives execution payload.

5. Workflow Manager access configuration database to get data about local jobs queue, available workers and their properties e.g.
    5.1. Workflow Manager via Calibration Manager retrieves the latest calibration data from the QPU, such as, coherence times, gate error rates and qubit frequencies by accessing Quantum Hardware Adapter. It could be realized either on demand or periodically.
    5.2. Latest calibration data with timestamp is stored in the configuration database.
6. Scheduling a hybrid job for execution on an available worker.
7. Accessing external service for optimization or other logic if there is a need.
8. Accessing hardware via adapter software.
    8.1. Accessing hardware adapter as part of separate use case from different worker.
9. Hardware adapter accesses QPU control stack via native API.
10. Results are returned to the adapter from QPU.

*A. Programming model explained*

Part of QCPaaS client SDK is RuntimeWorkerBase class outlined on the Fig.2. It is a base class that all custom user workloads should inherit from. It provides generic run() method that runtime looks for to execute scheduled workload. QMLWorker is an example implementation which contains an example hybrid algorithm that is planned to be executed on the runtime to increase efficiency. It overrides run() method and sets backend_name e.g. "FakeKolkata" and execution_options list e.g. ['ErrorMitigatedExecutionBackend', 'option2'] which is a pipeline (could be many items) for custom optimization logic that is present on the runtime and which is discovered as part of a separate process defined in section 1.2 of conceptual architecture vision description. RuntimeProvider (Fig.2) is used from the client to schedule workload on the remote runtime. Module name with a logic inherited from RuntimeWorkerBase is passed as a parameter (in our case qml_worker). Under the hood it converts python class to a string and discovers all required modules that are required to run the workload. This data is sent to the runtime via http call. On the runtime side Workflow Manager (Fig. 1) complements this programming model by looking for a worker that could satisfy workload needs in required packages, custom pipeline optimization modules.

One of the core drivers for QCPaaS development is need of a flexible programming model that allows to do pre and post-processing steps for different parts of algorithm execution, customise pipeline of quantum circuits transpilation, do optimisation with arbitrary custom routines or products from 3rd party vendors. Considering that development approach in the eco-system is classical Object-Oriented Programming (OOP) paradigm, the decision has been made to use classical design pattern from Object Oriented Design (OOD) the name for which is Decorator [33].

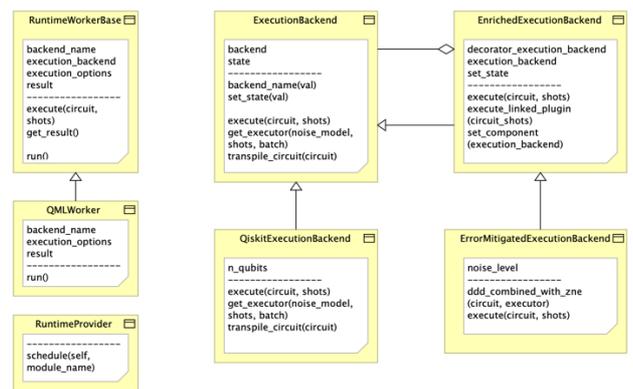

Fig. 2. Classes diagram for core part of the programming model.

It is structural design pattern that allows attaching new behaviors to objects by placing these objects inside special wrapper objects that contain the behaviors. This paradigm fits very nicely with logic of assigning pre and post-processing steps for the parts of algorithm execution in a reusable way. This could be achieved by developing custom pipeline classes e.g. ErrorMitigatedExecutionBackend and assigning correct order of their execution (pre, post) in execution_options list which comes as part of programming model. Reusability is achieved by deploying specific container images with the logic on the runtime using BYOC model which could be reused by different clients. On Fig.2 EnrichedExecutionBackend both aggregates ExecutionBackend and inherits from it. This a core of Decorator pattern, what allows to attach custom functionality (ErrorMitigatedExecutionBackend) on the fly to different backend implementations e.g QiskitBackend. Pattern itself doesn't handle full resolve and substitution of specific module under the hood of runtime. This is achieved by traversing all the descendants of EnrichedExecutionBackend which are available in the context of runtime execution and setting the correct order according to execution_options list.

*B. Hadrware adapters design*

This part of software usually goes with programming SDKs like Qiskit or Cirq. Qiskit has adapters architecture implemented via providers [10]. With a recent release of BackendV2 IBM team removed a lot of drawbacks and inflexibilities from previous version. e.g. hard to discover backend features due to lack of native data structures, handling of timing constraints. These limitations have been removed in version 2 along with new features being added like customizable compilation and comprehensive view of backend by representing multiqubit instructions. Of course, being opinionated and adjusted for IBM products line, still this part of Qiskit package could be used as a starting point when building new proxies for open hardware QPU back-ends. Abstraction should be created over Qiskit backend implementation, so substitute could be plugged-in (AD-6) in case licensing changes or compute optimization use case pop ups that will be stopping reaching optimization objectives (there were a couple in the previous version e.g. problems with mapping back-end operations, compilation customizability). Because of Qiskit widespread use, there are providers for main quantum hardware manufacturers e.g. IQM [12].

*C. Scheduling and availability mechanism*

Scheduling and availability mechanism works as a complementary part of Workflow Manager (Fig.1) and allows to schedule workloads for execution in a most efficient way. For utility, it requires careful consideration of the unique characteristics of quantum computing (QC), such as qubit coherence times, error rates, and the hybrid nature of quantum-classical computation. This paper doesn't go in details of optimization of these properties but tries to define important software modules that are required in QCPaaS as a baseline for optimization effort:
   a. Resource estimation. Based on which availability time is planned, as it allows to see how long algorithm execution could take. Microsoft built open-source version [14] which they use as part of their Azure quantum offering [13] and which could be incorporated in the implementation of this reference architecture. As explained in [16] it is quite untrivial task to equally distribute work across quantum machines. Their characteristics varies after day-to-day calibrations. Thus, different time-sensitive properties e.g. fidelity should be taking into consideration during scheduling.
   b. Queue management. Prioritization and scheduling of jobs based on various factors (size, estimated duration, user priority, SLAs etc.). In case batch job is scheduled (execution of 1000 and more iterations of optimization procedure) all of them should be executed within one scope to avoid long queue times when multiple users schedule their workloads.
   c. Reservation approach that allows full allocation of QH for a particular period.
   d. Policies that may include limitations on the number of concurrent jobs a user can submit.
   e. Feedback on the status of quantum jobs, including information on job progress, estimated completion times, and any issues encountered during execution, so jobs could be monitored and adjustments made based on feedback.

Most of the points mentioned above should be considered as optional, but important when trying to build enterprise level QCPaaS actively used by hundreds of users simultaneously. In the minimal implementation Job Scheduler should communicate with individual Hardware Adapters that share queue/availability information. Also, runtime should have simple queue management that allows scheduling jobs on hardware, performing retries in case of errors and handle priority access. This information could be used to build minimal functionality of "job will start in x time". Scheduling should be resilient to failures. There should be mechanism that handles cases when the provider session reaches timeout and continuation should be built to not lose results for long-running jobs e.g. 1 day and longer, what will be expensive to re-run.

What information could be retrieved from QH providers that could help with scheduling?
- IQM REST API [16] (used in Qiskit back-end). IQM has a very limited interface in terms of job visibility. They only provide information about the status of the job e.g. running, failing, waiting. No endpoints to control at least some aspects of the queue.
- Rigetti API [18]. Has reservation, which allows scheduling 15 minutes of priority access at the specified time. Cost optimization could be implemented with parallelization feature. On-demand access feature allows to run job without reservation, but it is not possible to estimate wait time because reserved jobs could take priority. For some reason there is no mechanism that updates wait time based on the new reservation job which is time fixed.

Examples above are provided with purpose to confirm that even established vendors don't standardize their core hardware APIs. Open QCPaaS initiative could also drive a standard for hardware API interface design, so emerging QH vendors and researchers could move faster with their new hardware platforms. In general, these differences in features and API designs create a lot of complexity and low-quality queue management/availability mechanism that are common on the market right now.

*D. Parametric compilation with calibration data*

Calibration data refers to the set of parameters and measurements (coherence times, qubit frequency, gate fidelities, error rates etc.) that describe the current physical state and performance characteristics of QPU. This data is essential for the effective use of QH, enabling more precise control over qubit behavior, reducing errors, and ultimately leading to more reliable quantum computations. Depending on the type of QH and many other external factors like temperature fluctuations, electromagnetic interference and many other noise sources calibration data could change many times per day. For long running quantum workloads, it is important to consider this. Quantum SDKs like Amazon Braket offer feature called "parametric compilation" [18] [20] as part of the hybrid jobs. Primarily it has been a mechanism to optimize computation by removing the need to compile circuit on every iteration of hybrid job what is quite time consuming. Along with that it also incorporates latest quantum device calibration data to ensure that results are produced with higher quality.

Intent of this section is an attempt to reverse engineer Amazon's implementation (considering that it is proprietary module hidden under AWS Braket platform) and outline architectural concerns that should be considering when implementing this feature in QCPaaS.

Main responsibility of Calibration Manager component on the Fig.1 is retrieval and storage of the latest calibration data from the quantum hardware. It could retrieve this data periodically or on demand. Assumption is that Hardware Adapters for QPU allows to get access to that data. Workflow Manager should ensure that execution is aligned with the real-time state of the hardware by using data from Calibration Manager. It should be smart enough to ensure minimal delay between compilation and execution to reduce the risk of hardware state changes. These two components rely heavily on the compiler that supports parametric compilation feature. As this reference architecture doesn't tie implementation to specific programming SDK which usually comes with compiler, this component is assumed to be chosen arbitrarily from the list of SDKs that support this feature e.g. Qiskit [20]. It will require some customization though. Functionality should be extended so that during the parameter binding phase, the retrieved calibration data is used to determine the optimal values for the circuit parameters. This could involve updating gate parameters like rotation angles based on qubit frequency or adjusting gate timings based on the latest coherence times. Also, there should be a change on the transpiler level to delay final compilation until the binding step is complete, ensuring that the calibration data is applied as close to execution time as possible. The transpiler would optimize the circuit based on this real-time data, potentially adjusting gate sequences or error mitigation strategies. Qiskit's execution pipeline should be modified to include a feedback loop where post-execution data (e.g., fidelity metrics) is used to inform future parameter bindings or calibration data updates. This could be quite complex endeavor and depends on main programming framework implementation. More detailed design could be planned once Calibration Management module of Qiskit [21] will be established. As of time of writing a paper it is in active development.

Some of the architectural concerns that should be covered during implementation include:

- Ensuring that the entire process—from retrieving calibration data to binding parameters and executing the circuit—happens swiftly is crucial to maintain accuracy.
- Quantum hardware is highly sensitive, and its properties can change rapidly. Ensuring that the compilation remains accurate in such an environment is a major challenge.
- Real-Time data integration: Integrating real-time calibration data into the compilation process without introducing significant delays requires efficient data management and processing pipelines.
- Complexity of Compilation: The Parametric compiler must be sophisticated enough to optimize the circuit based on detailed and potentially complex calibration data, which can involve intricate quantum mechanical considerations. As an alternative to Qiskit compiler, QIR [22] based implementation and it is benefits will be reviewed in the next sections.
- The system must be able to scale with the number of users and circuits being processed. This requires efficient management of calibration data and a robust scheduling system to handle potentially high volumes of parametric compilations.
- If the calibration data changes significantly between compilation and execution, or if the hardware state is not as expected, the system should either recompile the circuit or provide a mechanism for flagging or retrying the job.

A good candidate to incorporate into this design is implementation of open-source framework to perform quantum calibration and characterization is Qibocal [23]. Paper doesn't cover explicitly all the architectural concerns listed above but provides quite solid ground for moving forward with this important feature. Fresh look on the complexities of incorporating calibration data into error-aware compilation for NISQ devices is provided in [24]. Paper goes into an interesting direction of analysing historical calibration data and applying noise-aware compilation techniques based on it.

HYBRID QUANTUM-CLASSICAL COMPUTATIONS. QCPaaS AND OUTLOOK BEYOND.

QCPaaS optimizations gains that are coming as part of the design include:

- Reduced latency in circuits execution and in communication between classical compute and QPU.
- Parametric compilation that considers calibration data.
- Both efficient and effective scheduling mechanism.
- Batching iterations of long-running jobs together (as part of session mechanism).
- Flexible pipeline mechanism that allows to include custom optimization steps e.g. EM.

They provide benefit for algorithms like VQE and QAOA and offer what IBM calls [26] near-time execution benefit. Circuits initialization and communication between classical and quantum compute still takes significant time and prevent achieving real-time execution of classical code when maintaining QPU qubits coherence. To move closer to it, other areas of software stack should be covered as long as improvement in quantum hardware design.

### A. Needed improvements from the hardware side

As outlined in [27] classical computing has always been used in quantum for measuring results, control configurations etc. Microsoft quantum research group [25] puts emphasis on specialized hardware like arbitrary waveform generator (AWG) and field programmable gate array (FPGA) which are used in a common QPUs to handle precise pulses of microwave or lasers as part of the control stack. It amplifies that this specialized hardware comes with a loss of general-purpose compute features that are required for popular hybrid algorithms. Newer generations of quantum hardware are actively trying to address these challenges by introducing all required classical capabilities on a single chip. One prominent example is ultra-low-latency chip-to-chip links between quantum computers, GPUs and CPUs that is in active development by Seeqc [28]. The difficulties in effectively coordinating a dedicated accelerator with a central processing unit are not specific to quantum computing. Modern computing practices, especially the use of GPUs, have inspired approaches for managing data transfer between processors and improving code portability, as well as integration with existing tools and technologies. Little by little quantum is moving in that direction. Effort is led in collaboration with NVIDIA and their DGX Quantum platform [29]. DGX Quantum allows to deliver submicrosecond latency between GPU, QPU, accelerating hybrid workloads. Having an ability to deploy QCPaaS on DGX quantum and accessing QPU of choice from it (which should be physically close enough) would be very interesting set to explore new QU possibilities. Though, it doesn't look like NVIDIA is coming that way. At least not for now. But it also doesn't mean that Seeqc won't scale partnerships in broader community.

### B. Next steps on software stack level.

[1] defines OpenQASM as a main instrument for an aggressive quantum optimization research. It continues to being an important player, especially with OpenQASM 3.0 update which provides many features for real-time hybrid computations. It is a good choice for programming modern quantum-classical interactions. From another perspective, QIR [2] recommended itself to be more advantageous for optimizing and deploying quantum programs across platforms. QIR provides a powerful intermediate representation that is useful in large-scale quantum computing projects where performance optimization and cross-platform compatibility are critical. It typically requires a higher-level language to manage the real-time aspects of hybrid computation. It is adoption has been accelerated by Microsoft, Nvidia (which uses it as part of a programming model for CUDA-Q [31] and DGX Quantum respectively) and several other companies. From the observations, it feels more "native" and convenient for expressing the logic for data exchange and processing, while qubits remain live. QIR's foundation on LLVM IR [32] allows it to leverage the extensive LLVM toolchain for optimization and cross-compilation, including sophisticated analyses and transformations that are not directly applicable in OpenQASM. The seamless integration between quantum and classical code components that QIR aims to provide is more about the compilation and execution pipeline, including optimizations across the quantum- classical boundary. OpenQASM focuses on specifying the quantum circuit and its immediate classical control logic but does not inherently provide the same level of support for optimization and cross- compilation as QIR does. [25] work focus is on innovative software components that leverage QIR to remove limitations of existing established compute models that raise efficiency of quantum algorithms. Results look promising, so there is a very solid ground to continue further research based on QIR.

With all the innovation expected in both hardware and software stacks, QCPaaS will continue to serve a role of a mandatary platform in overall eco-system, adopting it is internal design along with programming model to the needs of programming stacks based on QIR and OpenQASM. As of now they could be fully integrated and are first class citizens in QCPaaS architecture vision. Container with programming SDK based on QIR e.g. NVIDIA-Q is deployed to the runtime. End-user designs and algorithm using the same SDK and schedules workload for execution, for which runtime identifies container with NVIDIA-Q. Physically, classical-quantum compute is co-located via hypothetical low link connections to bring full-potential for hybrid algorithms execution. If assumption on the eco-system development is correct – QU potential looks close enough.

### Discussion

As mentioned in [1] industry effort focus is primarily on cloud-based quantum hardware exposure and utilization along with tooling that helps to design for algorithmic advancements. This reflects profit-driven capitalism, which is normal considering that significant investments done in QC capabilities of these companies should be pay offed in at least mid-term with minimal risks. In the center of this value chain there are major inefficiencies caused by closed ecosystem implementations of runtime environments and inflexible programming models. There is a lot of room for improvement of QC execution quality which could be done using the available tooling, but our hypothesis is that by opening the design of QCPaaS, allowing everyone to contribute and customize it, could have a profound effect and major push towards reaching QU in case standard will be actively adopted. We want to believe that it will also stimulate IBM, AWS, Microsoft and major hardware vendors to open-source their quantum runtimes, so the best of the breed standardized runtime platform could emerge in the upcoming years and industry wide one more milestone could be reached.

Paper complements QCOM architecture [1] by providing required tooling for implementation of optimization steps that require remote/co-located execution.

QCPaaS reference architecture defined as part of this research plays a confident and irreplaceable role in near-term future of QU. While more in-depth analysis & prototyping activities are required to cover complexities and unknowns in scheduling, hardware adapters mechanism etc. most of the architectural decisions will apply to a most of remote execution use cases.

Next steps in research should be taking a prototype of the architecture and deploying it close to real QPU hardware implementation to measure quantitative metrics e.g. fidelity. Scheduling mechanism, parametric compilation implementation along with a proposal for hardware adapters mechanism could be a nice vitamin and improvement for the QCPaaS. Priority of showing how to leverage proposed programming model flexibility for QC quality improvement could be outlined as one important direction. From another perspective refining usage of QIR as part of this software stack for real-time hybrid compute scenarios shouldn't be skipped, considering it growing adoption in the community.

### Conclusion

This study demonstrates the importance of standardization and adoption of ecosystem-agnostic QCPaaS architecture in the aggressive QC optimization research. Value proposition driven by industry needs transition into a proposed QCPaaS architecture and design of its core components and programming model. Role of the hybrid QC is outlined along with improvements that are required on both software and hardware stacks to achieve real-time hybrid QC.